\documentclass[aps,prapplied,reprint,amsmath,amssymb]{revtex4-2}
\usepackage{graphicx}
\usepackage{siunitx}
\usepackage{booktabs}
\usepackage{array}
\usepackage{amsmath,amssymb}
\usepackage{amsmath}
\usepackage{bm}
\usepackage{booktabs}
\usepackage[colorlinks=true,linkcolor=blue,citecolor=blue,urlcolor=blue]{hyperref}
\usepackage[caption=false]{subfig}
\usepackage{placeins}
\usepackage{breakcites}

\begin{document}

\title{Depletion-limited Effective Hall mobility in Micrometer-Scale High-Purity Germanium Crystals}

\author{Narayan Budhathoki}
\affiliation{Department of Physics, University of South Dakota, Vermillion, SD 57069, USA}
\author{Dongming Mei}
\email{Dongming.Mei@usd.edu}
\affiliation{Department of Physics, University of South Dakota, Vermillion, SD 57069, USA}
\author{Sanjay Bhattarai}
\affiliation{Department of Physics, University of South Dakota, Vermillion, SD 57069, USA}
\author{Sunil Chhetri}
\affiliation{Department of Physics, University of South Dakota, Vermillion, SD 57069, USA}
\author{Kunming Dong}
\affiliation{Department of Physics, University of South Dakota, Vermillion, SD 57069, USA}
\author{Shasika Panamaldeniya}
\affiliation{Department of Physics, University of South Dakota, Vermillion, SD 57069, USA}
\author{Athul Prem}
\affiliation{Department of Physics, University of South Dakota, Vermillion, SD 57069, USA}
\author{Austin Warren}
\affiliation{Department of Physics, University of South Dakota, Vermillion, SD 57069, USA}

\date{\today}

\begin{abstract}
Electrostatic effects can strongly constrain charge transport in thinned high-purity germanium (HPGe), with direct implications for radiation detectors and Ge-based electronic and quantum devices. We report a systematic experimental characterization of the thickness-dependent effective Hall mobility in bulk-grown, detector-grade HPGe at room temperature using Hall-effect measurements on n- and p-type samples sequentially thinned from 2.7~mm to 7~\textmu m. The intrinsic bulk carrier mobility remains thickness independent in this regime; the observed reduction in Hall-extracted mobility arises from electrostatic surface depletion that reduces the electrically active conducting thickness. The thickness-dependent data are accurately parameterized by an empirical extended-exponential relation, $\mu(t)=\mu_{0}[1-\exp(-(t/\tau)^{\beta})]$, where $\tau$ is a characteristic electrostatic length scale. Comparison with boundary-scattering and depletion-based models shows that Fuchs--Sondheimer scattering is negligible, while electrostatic depletion dominates the transport behavior. The hierarchy $\lambda_{D}<\tau\lesssim W_{0}$ directly links the apparent mobility reduction to long-range screening and near-surface electric fields.
These results yield a simple design guideline: maintaining thicknesses $t\gtrsim 3\tau$ preserves near-bulk transport, whereas thinner structures operate in a depletion-controlled regime with strongly reduced effective conductivity.
\end{abstract}
\maketitle
\section{Introduction}
High-purity germanium (HPGe) remains a benchmark semiconductor for radiation detection, infrared photonics, and emerging quantum technologies due to its exceptional crystalline purity, high carrier mobility, and narrow band gap~\cite{DARKEN199523,knoll2010radiation,gilmore2008practical,nakano2020ultra}. These intrinsic material properties enable efficient charge collection, low electronic noise, and excellent energy resolution over a wide range of operating conditions, making HPGe indispensable for low-threshold ionization detectors and precision semiconductor devices~\cite{mei2025phonon,mei2025quantum_sensors,budhathoki2025aps_irfiber}.

Beyond detector applications, germanium has re-emerged as a promising channel material for next-generation microelectronics as silicon complementary metal-oxide-semiconductor (CMOS) technology approaches fundamental electrostatic and mobility scaling limits~\cite{mosfet1,mosfet0,mosfet2,mosfet4,mosfet5,GeoI,i}. Among group-IV semiconductors, Ge exhibits the highest hole mobility ($\mu_h \approx 1900~\mathrm{cm^2/V.s}$)~\cite{Myronov2014JJAP,jacoboni1983electronic} and an electron mobility well above that of silicon~\cite{mosfet6,mosfet7,mosfet8,mosfet9,mosfetandmob,mosfetandmobilitydeg}, motivating continued development of Ge-on-insulator (GeOI) platforms and ultra-thin-body Ge transistors.

In both detector and device contexts, Ge structures are increasingly thinned to micrometer or sub-micrometer dimensions to enable faster charge collection, improved electrostatic control, phonon engineering, and hybrid integration. Importantly, in the micrometer thickness regime relevant to bulk-grown HPGe, the intrinsic carrier mobility is not expected to depend on thickness, as phonon and impurity scattering dominate momentum relaxation and the mean free path remains orders of magnitude smaller than the physical dimensions. Nevertheless, electrical transport measurements—particularly Hall-effect measurements—can exhibit a pronounced apparent thickness dependence when electrostatic surface depletion reduces the electrically active conducting cross-section.

In such cases, the experimentally extracted Hall mobility represents an effective Hall mobility, reflecting both the intrinsic bulk mobility and the fraction of the sample thickness that remains electrically neutral and conducting. As the physical thickness approaches characteristic electrostatic length scales, such as the Debye screening length and surface depletion width, long-range surface electric fields can significantly compress the conducting channel without modifying the local microscopic scattering mechanisms. A clear and quantitative understanding of how electrostatic depletion constrains the effective conducting thickness in thinned bulk HPGe is therefore essential for correctly interpreting transport measurements and for guiding the design of high-performance detectors, Ge-based electronic, and quantum devices.
In such devices—and in thinned HPGe detectors—the electrically active channel thickness is often reduced to tens of micrometers or below, making it essential to understand how electrostatic effects constrain charge transport as a function of thickness, rather than implying a modification of intrinsic carrier mobility.

Carrier transport in thin semiconductor structures can be influenced by several mechanisms, including surface-field-induced depletion, long-range Coulomb interactions, interface disorder, and strain-induced band-structure modification~\cite{fischetti1996band,nakano2016high}. 
While these effects have been extensively investigated in Si, strained-Ge, and GeOI technologies~\cite{ando1982electronic,fischetti1996band,mosfetandmobilitydeg}, a systematic experimental framework characterizing how electrostatic depletion governs effective charge-carrier mobility in high-purity bulk-grown Ge remains largely unexplored. In particular, the interplay between electrostatic screening, surface depletion width, and intrinsic band anisotropy has not been quantitatively established for mechanically thinned HPGe crystals, where interface-induced scattering is minimal and electrostatic effects are expected to dominate.

Prior studies of transport in Ge and related systems have largely focused on Ge-on-insulator (GeOI) and ultrathin epitaxial films, where conduction is strongly influenced by interface disorder, strain relaxation, and gate-stack engineering~\cite{GeoI,mosfetandmobilitydeg,nakano2016high}. These works provide valuable insight into technologically relevant metal–oxide–semiconductor (MOS) channel structures, but they do not address bulk-grown, detector-grade Ge that is mechanically thinned to the few-micrometer scale while preserving ultra-low impurity concentrations and minimal interface damage. In contrast, the present work reports the first systematic investigation into the thickness-dependent effective Hall mobility of bulk high-purity Ge, covering a continuous range from $2.7~\mathrm{mm}$ down to $7~\mu\mathrm{m}$ on the same material platform. 
This geometry is directly relevant to HPGe detectors, Ge-based quantum sensors, and Ge electronics in which devices are thinned for charge-collection speed, phonon engineering, or hybrid integration, yet must retain near-bulk transport properties~\cite{mei2025phonon,mei2025quantum_sensors,mosfet1}.

Classical boundary-scattering models such as the Fuchs--Sondheimer (FS) formalism predict only weak corrections to transport when the carrier mean free path is much smaller than the physical thickness, as is the case for HPGe at room temperature~\cite{Fuchs1938,sondheimer1952mean}. 
In contrast, electrostatic surface depletion can substantially reduce the electrically active conducting thickness when the physical thickness approaches the Debye screening length, leading to a pronounced reduction in the Hall-extracted effective mobility without altering the intrinsic bulk scattering mechanisms~\cite{fischetti1996band}. 
A compact empirical description that captures this electrostatically governed transition from bulk-like to depletion-limited conduction is therefore valuable for both detector optimization and the design of advanced Ge-based electronic and quantum devices.

In this work, we experimentally investigate effective Hall mobility in USD-grown HPGe crystals fabricated at the University of South Dakota. 
Van der Pauw Hall-effect measurements were performed on p - and n -type samples sequentially thinned from $2.7~\mathrm{mm}$ to $7~\mu\mathrm{m}$. 
A central result is that the the thickness-dependent effective Hall mobilities are accurately
described by an empirical extended-exponential relation,
\begin{equation}
\mu(t)=\mu_0\!\left[1-\exp\!\bigl(-(t/\tau)^\beta\bigr)\right],
\label{eq:extended-exponential}
\end{equation}
where $\mu_0$ denotes the intrinsic bulk mobility, $\tau$ represents a characteristic electrostatic length scale, governing the reduction of the conducting channel, and $\beta$ is a phenomenological shape parameter.
Comparison with FS-only and depletion-only models shows that boundary scattering is negligible, whereas surface-field-induced depletion governs the observed the thickness dependent effective Hall mobility.
The extracted characteristic lengths satisfy the hierarchy $\lambda_D < \tau \lesssim W_0$, directly linking the apparent mobility reduction to long-range screening and near-surface electric fields rather than to geometric confinement. This connection provides a simple device-level design rule: maintaining thicknesses $t \gtrsim 3\tau$ preserves near-bulk transport, whereas structures with $t \lesssim \tau$ operate in a depletion-controlled regime with strongly reduced electrically active conduction. In this work, all measurements are performed at room temperature
($T = 295~\mathrm{K}$), and the conclusions presented here are restricted to
transport behavior in this temperature regime.

Overall, this study establishes a consistent electrostatic framework for interpreting thickness-dependent effective Hall transport in HPGe, providing quantitative guidance for optimizing high-resolution HPGe detectors and designing thin-body Ge-based electronic and quantum devices in which electrostatic control must be carefully balanced against transport performance.

\section{Methods}

\subsection{Crystal Growth and Material Preparation}

High-purity germanium crystals were grown in-house using successive zone refining and Czochralski pulling. Commercial Ge ingots with initial impurity concentrations of $(10^{13}$–$10^{14})~\mathrm{cm^{-3}}$ were refined to $(2$–$3)\times10^{11}~\mathrm{cm^{-3}}$~\cite{bhattarai2024crystals}. 
Prior to melting, all components contacting the charge were cleaned in DI water and acetone, followed by an HNO$_3$:HF (3:1) etch for Ge surfaces and H$_2$O$_2$ treatment for quartz hardware. The charge was melted in a quartz crucible on a graphite susceptor using RF induction heating. Chamber pressure was reduced to $20~\mathrm{Pa}$ and subsequently to $\sim 2\times10^{-3}~\mathrm{Pa}$ before stabilization.

Crystal growth proceeded from a $\langle100\rangle$ seed. Boules (8–8.5~cm diameter) were pulled at $40~\mathrm{mm/h}$ during necking and $22~\mathrm{mm/h}$ during body growth in 5N H$_2$. An initial Dash neck suppressed dislocations, followed by constant-diameter growth and a tapered tail-off~\cite{dash1958method,bhattarai2024crystals}. 
The crystals were cooled over a 10~h controlled cycle and removed for characterization.

\subsection{Sample Thinning and Surface Preparation}

Samples were thinned using sequential mechanical and chemical processing. Wafers were first sliced using a diamond wire saw and mechanically lapped to approximately $300~\mu\mathrm{m}$. Coarse and fine planarization were performed with 10~µm and 5~µm Al$_2$O$_3$ abrasives to achieve a uniform surface.

\begin{figure}[ht!]
\centering
\includegraphics[width=0.33\textwidth]{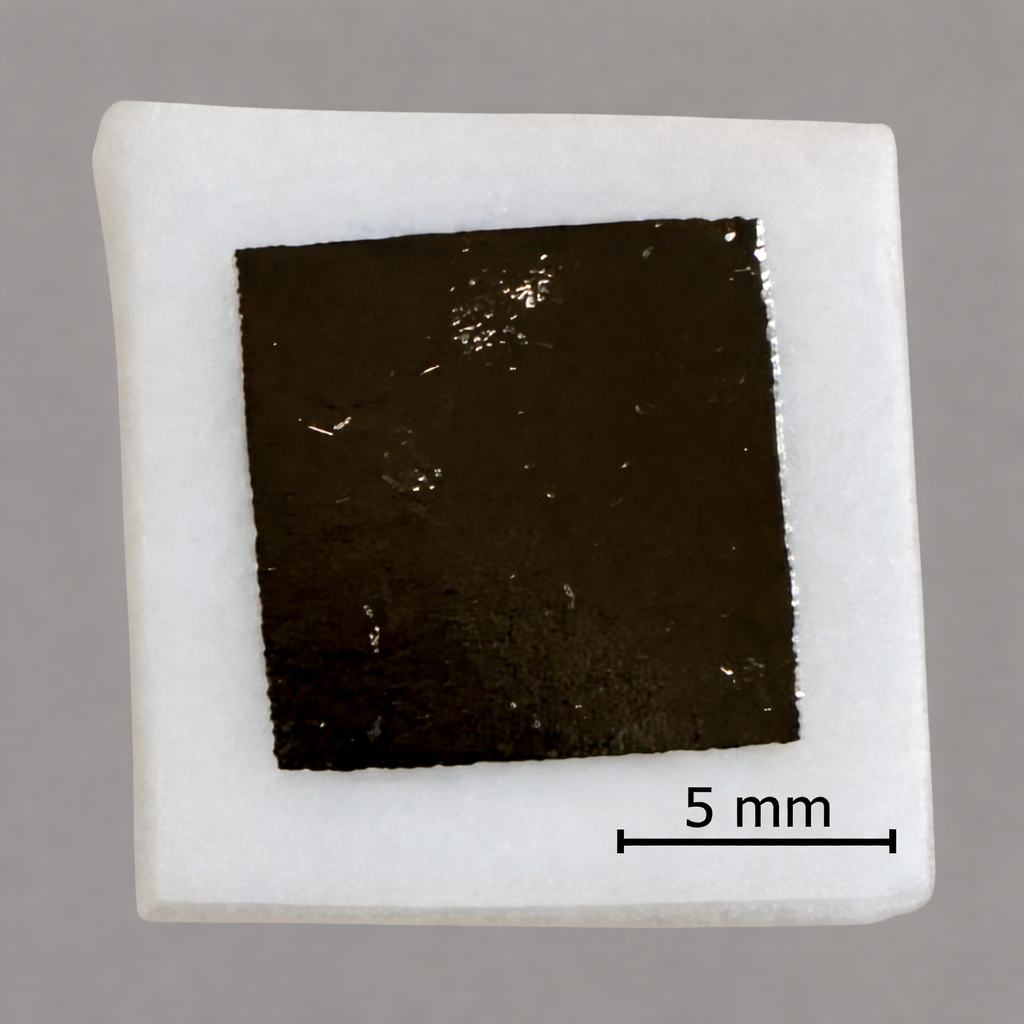}
\caption{Photograph of a representative HPGe sample mounted on a PTFE holder
during thinning and Hall measurements. The Ge crystal has lateral dimensions
of approximately $10 \times 10~\mathrm{mm}^2$.
The scale bar corresponds to $5~\mathrm{mm}$.}
\label{fig:sample_photo}
\end{figure}

Final thickness control was obtained by iterative wet chemical etching in HF:HNO$_3$ (1:6), which provides uniform material removal with minimal subsurface damage for HPGe~\cite{bhattarai2024crystals}. During etching, each sample was mounted on a PTFE base using Crystalbond wax, which provided mechanical support and resistance to HF. A representative mounting configuration is shown in Fig.~\ref{fig:sample_photo}.

After each etching cycle, samples were rinsed in DI water, dried under nitrogen, and demounted for inspection. Optical microscopy verified the absence of pits or etch-induced defects at the resolution limit, and stylus profilometry confirmed the thickness with a typical uncertainty of $\pm 1~\mu\mathrm{m}$. This protocol reproducibly produced samples ranging from the bulk scale $2.7~\mathrm{mm}$ down to $7~\mu\mathrm{m}$ without detectable surface deterioration relative to as-grown detector-grade material~\cite{bhattarai2024crystals}.
\subsection{Thickness Determination}

The thickness of each sample was determined using the mass--density--area method. 
Prior to each thinning step, the sample was cut to a known area 
($A \approx 1~\mathrm{cm^2}$). After each controlled acid etch, the sample was 
rinsed, dried, and the remaining mass ($m$) was measured using a 
0.1~mg precision balance. Using the density of Ge 
($\rho = 5.323~\mathrm{g/cm^3}$), the instantaneous thickness after each etch 
was computed as:
\begin{equation}
 t = \frac{m}{\rho A}.
\end{equation}
This approach provided sub-micrometer accuracy throughout the sequential thinning 
process and was cross-validated against profilometry measurements.
\subsection{Electrical Transport Measurements}

\begin{figure}[ht]
\centering
\includegraphics[width=0.75\linewidth]{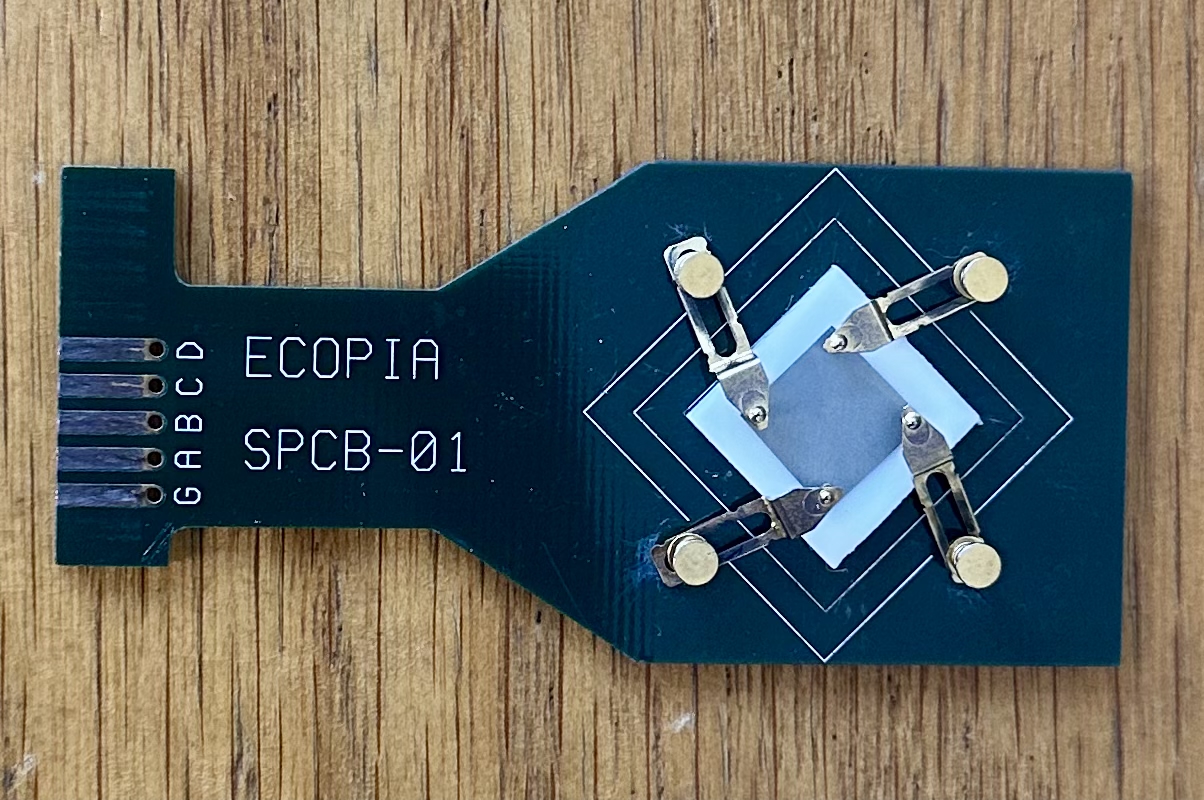}
\caption{HPGe sample mounted in a Van der Pauw geometry with four indium corner contacts.}
\label{fig:vdp_setup}
\end{figure}

Room-temperature ($295~\mathrm{K}$) electrical transport measurements were performed using the Van der Pauw Hall-effect technique on an Ecopia HMS-3000 system. Four indium ohmic contacts were placed at the corners of each HPGe specimen to ensure uniform current injection and minimal contact resistance. A representative sample mounted on the carrier board is shown in Fig.~\ref{fig:vdp_setup}.

\begin{figure}[ht]
\centering
\includegraphics[width=0.85\linewidth]{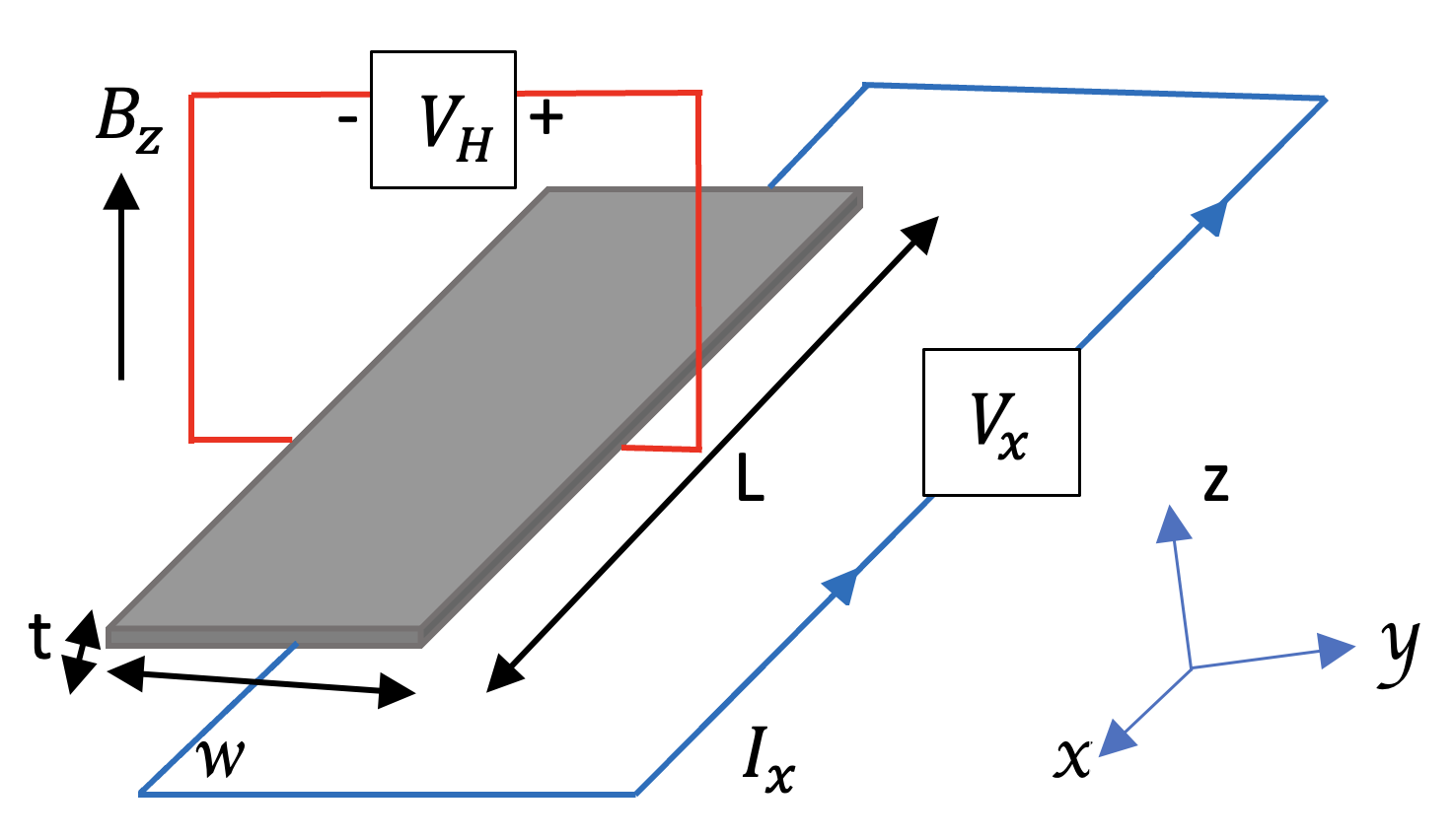}
\caption{Hall-effect measurement geometry. A current $I_x$ flows along the $x$-direction under a perpendicular magnetic field $B_z$, producing a transverse Hall voltage $V_H$.}
\label{fig:hall_scheme}
\end{figure}

During each measurement, a constant excitation current in the range $1$–$200~\mu\mathrm{A}$ was applied while a perpendicular magnetic field up to $0.58~\mathrm{T}$ was swept. The longitudinal voltage $V_x$ and the transverse Hall voltage $V_H$ were recorded simultaneously, enabling the determination of resistivity, carrier concentration, and Hall mobility. The measurement configuration is illustrated schematically in Fig.~\ref{fig:hall_scheme}.

Ohmic behavior and the absence of self-heating were verified by confirming that $V_x$ varied linearly with $I_x$ and that the extracted transport parameters were independent of excitation current within experimental uncertainty over the full current range. The resistivity is given by
\begin{equation}
\rho = R_s\, t,
\end{equation}
where $R_s$ is the sheet resistance obtained from the Van der Pauw relation and $t$ is the measured sample thickness. The Hall voltage satisfies~\cite{hall}
\begin{equation}
V_H = -\,\frac{I_x B_z}{n q\, t},
\label{eq:VH}
\end{equation}
where $q$ is the elementary charge and $n$ is the carrier concentration. The Hall mobility is then obtained from~\cite{mu}
\begin{equation}
\mu = \frac{1}{q n \rho}
    = \frac{|V_H|\, t}{I_x B_z\, \rho},
\label{eq:mu}
\end{equation}
with the corresponding carrier concentration given by~\cite{mu}
\begin{equation}
n = -\,\frac{I_x B_z}{q\, t\, V_H}.
\label{eq:n}
\end{equation}

We emphasize that the Hall mobility extracted using Eq.~(\ref{eq:mu}) represents an effective Hall mobility, evaluated under the standard assumption that current flows uniformly across the full geometric thickness $t$. In the presence of electrostatic surface depletion, the electrically active conducting thickness can be smaller than $t$, while the intrinsic bulk mobility within the neutral region remains unchanged. Consequently, any observed thickness-dependent effective Hall mobility reflects a geometric and electrostatic reduction of the conducting cross-section rather than a modification of microscopic carrier scattering mechanisms.

The overall uncertainty in the extracted transport parameters is within $\pm 1 \%$, dominated by geometric variations in contact placement and instrument precision. Each reported mobility value represents the average of at least three independent measurement runs performed at both magnetic-field polarities to ensure reproducibility.

\subsection{Electrostatic depletion and effective conducting thickness}

Before introducing the mobility modeling, it is essential to clarify the
geometric and electrostatic origin of the thickness dependence observed in the
Hall-extracted mobility. Although carrier mobility in a homogeneous bulk
semiconductor is an intrinsic material property governed by phonon and impurity
scattering, thinning a high-purity Ge crystal introduces electrostatic surface
effects that modify the electrically active conducting cross-section without
altering the local microscopic mobility.

\begin{figure}[t]
    \centering
    \includegraphics[width=0.48\textwidth]{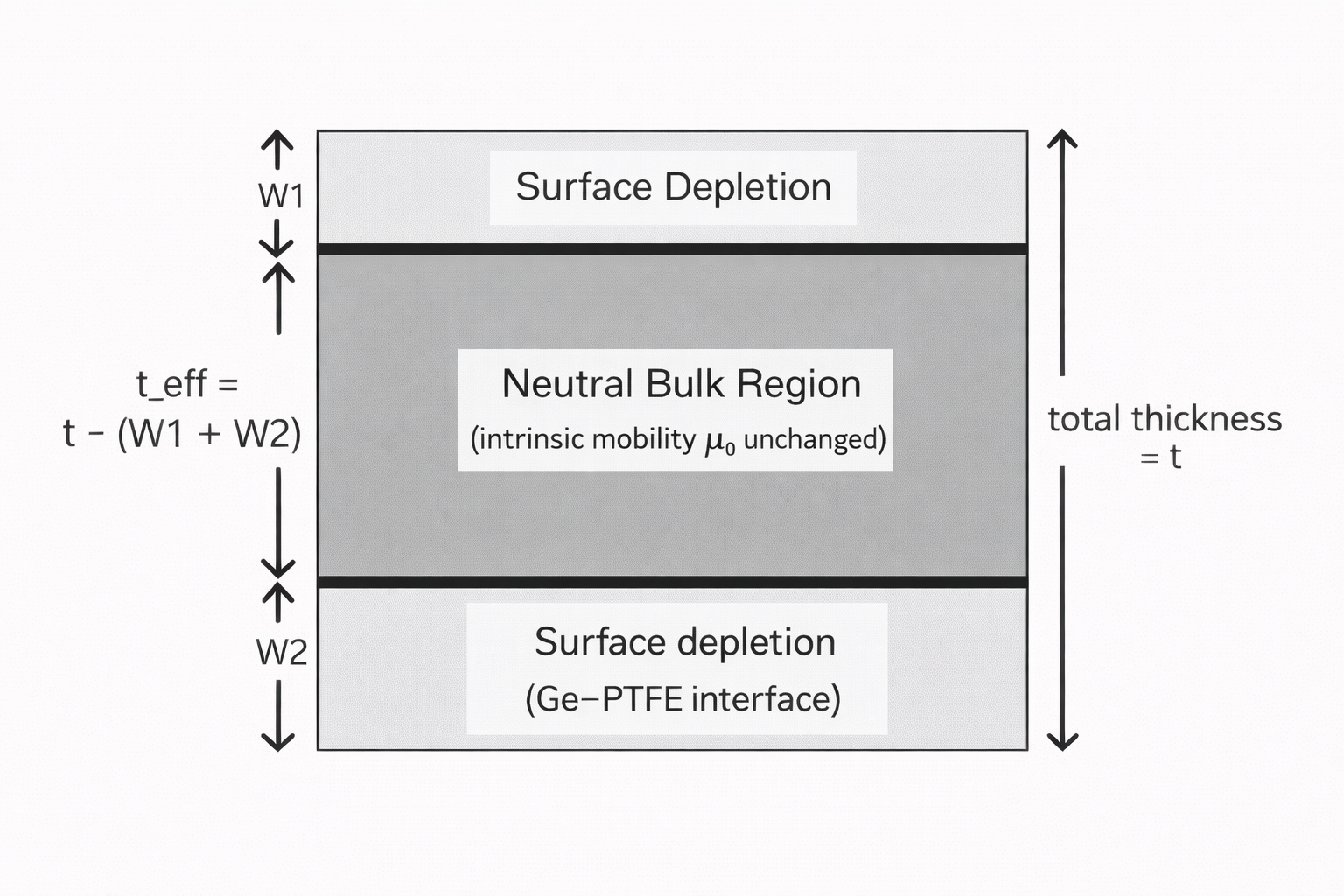}
    \caption{Schematic cross section of a thinned high-purity germanium (HPGe)
    sample illustrating electrostatic surface depletion. Band bending at the
    top free surface and at the Ge--PTFE interface produces depletion regions
    of widths $W_1$ and $W_2$, respectively. The electrically active neutral
    region has an effective thickness
    $t_{\mathrm{eff}} = t - (W_1 + W_2)$. The intrinsic carrier mobility
    $\mu_0$ within the neutral bulk region remains unchanged; the reduction in
    Hall-extracted mobility arises from the reduced conducting cross section
    assumed in the Hall analysis.}
    \label{fig:schematic}
\end{figure}

Figure~\ref{fig:schematic} schematically illustrates this effect for the present
HPGe samples. Surface band bending at the top free surface and at the Ge--PTFE
interface produces depletion regions of widths $W_1$ and $W_2$, respectively.
As a result, the electrically active neutral region has an effective thickness
\begin{equation}
t_{\mathrm{eff}} = t - (W_1 + W_2),
\end{equation}
where $t$ is the total physical thickness. Within this neutral bulk region, the
intrinsic carrier mobility $\mu_0$ remains bulk-like and thickness independent.

In contrast, the Hall-effect analysis assumes the full geometric thickness $t$
when converting the measured voltages into resistivity, carrier density, and
mobility. Consequently, as the depletion regions expand with decreasing sample
thickness, the Hall-extracted mobility is reduced according to
\begin{equation}
\mu_{\mathrm{Hall}} = \mu_0\,\frac{t_{\mathrm{eff}}}{t},
\end{equation}
even though the intrinsic bulk mobility $\mu_0$ remains unchanged~\cite{hall}. The observed
thickness dependence, therefore, reflects a geometric and electrostatic effect
associated with surface depletion, rather than a modification of microscopic
scattering mechanisms or the emergence of a new transport regime. With this distinction between intrinsic mobility and effective Hall mobility
established, we now introduce an empirical model to parameterize how the
electrically active conducting thickness evolves with the total sample thickness
under electrostatic depletion.

\subsection{Modeling and Data Analysis}

The experimentally measured Hall mobilities were analyzed using the extended-exponential relation introduced in Eq.~\eqref{eq:extended-exponential}. We emphasize that this functional form is employed as an empirical parameterization of how the electrically active conducting fraction evolves with sample thickness under electrostatic surface depletion, rather than as a microscopic transport law. In this framework, the intrinsic bulk mobility $\mu_0$ is treated as thickness independent, while deviations from the bulk limit arise from a reduction of the conducting channel.

The characteristic length scale $\tau$ extracted from the fits was compared with two electrostatic length scales: the Debye screening length $\lambda_D$ and the zero-temperature depletion width $W_0$. The Debye length is given by~\cite{debye}
\begin{equation}
\lambda_D = \sqrt{\frac{\varepsilon k_B T}{q^2 N_{\mathrm{eff}}}},
\end{equation}
where $N_{\mathrm{eff}}$ is the Hall-extracted effective impurity concentration. The depletion width $W_0$ was obtained from the electrostatic model described in Sec.~\ref{sec:results}. These comparisons provide a quantitative basis for linking the fitted parameter $\tau$ to physically meaningful electrostatic length scales governing the penetration of surface electric fields into the bulk.

To benchmark the empirical description, the fitted results were compared with two classical limiting cases:  
(i) the Fuchs--Sondheimer (FS) surface-scattering model, and  
(ii) a depletion-only electrostatic model in which the electrically active conducting thickness depends explicitly on the surface potential.  
Consistent with expectations for room-temperature HPGe, FS scattering predicts only negligible thickness dependence due to the short phonon-limited mean free path, while the depletion-only model captures the dominant experimental trend and provides physical bounds for the fitted values of $\tau$. The extended-exponential form offers a compact parameterization that interpolates smoothly between the bulk-like and depletion-limited regimes without invoking additional microscopic scattering mechanisms.

All data analysis and visualizations were performed using \texttt{Matplotlib}~\cite{matplotlib}. Uncertainties were propagated using standard Gaussian error propagation, and uncertainties in the fitted parameters were obtained from the covariance matrix returned by the nonlinear fitting routine.

\section{Results}
\label{sec:results}
\subsection{Characteristic Electrostatic and Transport Scales}

\begin{figure}[htbp]
\centering
\includegraphics[width=0.50\textwidth]{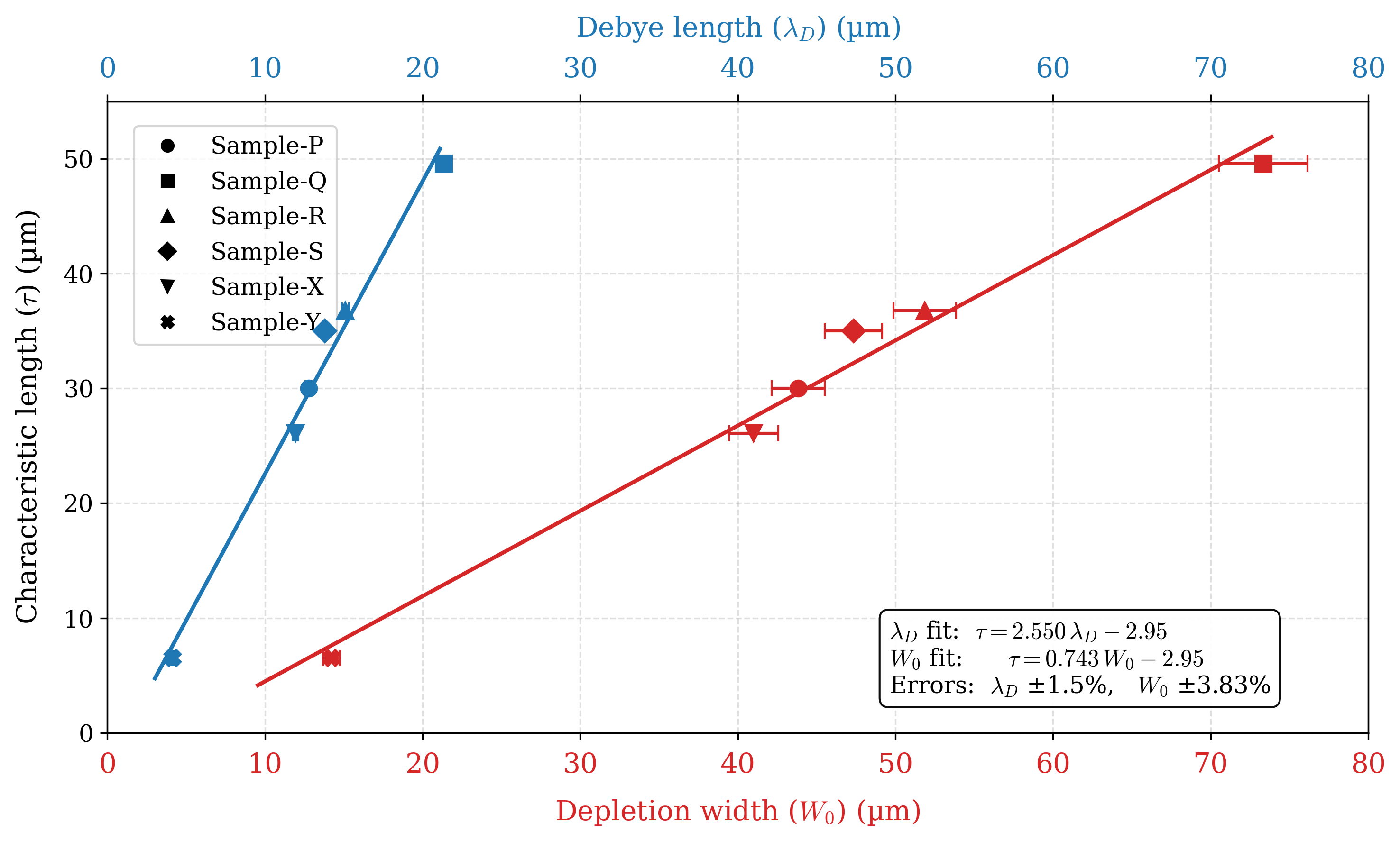}
\caption{Electrostatic and transport length scales for USD-grown HPGe samples (P, Q, R, S, X, and Y). Symbols denote the Debye length ($\lambda_D$), depletion width ($W_0$), and fitted electrostatic characteristic length ($\tau$). Linear fits quantify how $\tau$ tracks the electrostatic length scales across all samples.}
\label{fig:lengths}
\end{figure}

The characteristic electrostatic and transport length scales governing the thickness-dependent effective Hall mobility, arising from electrostatic reduction of the conducting thickness in HPGe are summarized in Table~\ref{tab:params-compact} and plotted in Fig.~\ref{fig:lengths} for all USD-grown samples (P, Q, R, S, X, and Y). Together, these scales establish a consistent physical hierarchy that constrains how the electrically active conducting thickness evolves as the crystal is thinned under electrostatic surface depletion.

\begin{table*}[t]
\centering
\fontsize{7}{15}\selectfont
\caption{Characteristic electrostatic and transport scales for HPGe at $T=295$~K. Parameters are representative of the experimental samples shown in Fig.~\ref{fig:lengths}.}
\label{tab:params-compact}
\begin{ruledtabular}
\begin{tabular}{lcccc}
Model/ Scale &
Governing expression &
Representative inputs &
Value/ Interpretation \\
\hline
Debye length~\cite{debye} &
$\lambda_D=\sqrt{\frac{\varepsilon k_B T}{q^2 n}}$ &
$\varepsilon_r=16.2$, $T=295$~K, $n=(5$--$133)\times10^{10}$~cm$^{-3}$ &
$\lambda_D=4$--$21~\mu$m (decreases with $N_{\mathrm{eff}}$) \\

Surface depletion width~\cite{mu} &
$W_0=\sqrt{\frac{2\varepsilon(V_{bi}-V)}{qN}}$ &
\emph{Assumed} $(V_{bi}-V)=0.1$--$0.3$~V; $N=(5$--$133)\times10^{10}$~cm$^{-3}$ &
$W_0=14$--$73~\mu$m; always $>\lambda_D$ \\

Mean free path~\cite{mfp} &
$\lambda_{\mathrm{mfp}}=\frac{\mu_0 m^* v_{th}}{q}$ &
$\mu_0=1900~\mathrm{cm^2/(V\cdot s)}$; $m^*_e=0.12\,m_0$ &
$\lambda_{\mathrm{mfp}}\approx44$~nm (phonon-limited) \\

Empirical mobility fit (h$^+$) &
$\mu(t)=\mu_0\!\left[1-e^{-(t/\tau)^{\beta}}\right]$ &
$\mu_0=1899.1~\mathrm{cm^2/(V\cdot s)}$; $\tau=40.4~\mu$m; $\beta=1.12$ &
$\tau>\lambda_D$ \\

Empirical mobility fit (e$^-$) &
$\mu(t)=\mu_0\!\left[1-e^{-(t/\tau)^{\beta}}\right]$ &
$\mu_0=2349.2~\mathrm{cm^2/(V\cdot s)}$; $\tau=24.6~\mu$m; $\beta=0.59$ &
$\tau>\lambda_D$ \\

FS Scattering~\cite{Fuchs1938} &
$\mu(t)\approx\mu_0\!\left[1-\frac{3\lambda_{\mathrm{mfp}}}{8t}(1-p)\right]$ &
$p\in[0,1]$ (specularity factor) &
Negligible for $t>10~\mu$m \\

\end{tabular}
\end{ruledtabular}

\normalsize
\end{table*}

The Debye screening length $\lambda_D$ ranges from $4$–$21~\mu$m across the
measured impurity concentrations $(5$–$133)\times10^{10}~\mathrm{cm^{-3}}$,
decreasing systematically with increasing doping. The surface depletion width
$W_0$ spans $14$–$73~\mu$m over the same range, based on an \emph{assumed}
surface band bending of $0.1$–$0.3$~V, consistent with literature-reported
values for Ge surfaces exposed to air or dielectric interfaces~\cite{debye}. This surface
potential is not directly measured in the present work and is used solely to
estimate the relevant electrostatic length scales. For all samples,
$W_0>\lambda_D$, indicating that surface electric fields penetrate deeply into
the crystal and can substantially reduce the electrically active conducting
thickness.

By contrast, the phonon-limited mean free path is only $\lambda_{\mathrm{mfp}}\approx44~\mathrm{nm}$, more than three orders of magnitude smaller than the thinnest specimen. Consequently, diffuse boundary scattering is insufficient to explain the observed reduction in the effective Hall mobility; the condition $t/\lambda_{\mathrm{mfp}}\gg100$ holds for all samples, placing the system firmly in the bulk-scattering regime.

Across the full set of samples, the fitted characteristic length $\tau$ obeys the hierarchy
\begin{equation}
\lambda_{\mathrm{mfp}} \ll \lambda_D < \tau \lesssim W_0,
\label{eq:hierarchy}
\end{equation}
demonstrating that the observed thickness dependence of the Hall-extracted effective mobility reflects electrostatic constraints on the conducting thickness rather than any modification of the intrinsic carrier mobility or boundary scattering. A central result of Fig.~\ref{fig:lengths} is that $\tau$ exhibits approximately linear correlations with both $\lambda_D$ and $W_0$.

\paragraph*{a. Correlation between $\tau$ and $\lambda_D$:}
A global linear fit yields
\[
\frac{d\tau}{d\lambda_D} \approx 2.55,
\]
indicating that $\tau$ increases more than twice as rapidly as the Debye length. This behavior reflects the dominant role of long-range electrostatic screening: as $\lambda_D$ increases under weaker doping, surface electric fields penetrate further into the bulk, allowing the electrically active conducting region to expand accordingly. The slope greater than unity suggests that $\tau$ is influenced not only by screening but also by the gradual relaxation of the depletion potential.

\paragraph*{b. Correlation between $\tau$ and $W_0$:}
A second linear trend is observed between $\tau$ and the depletion width, with a smaller slope,
\[
\frac{d\tau}{dW_0} \approx 0.74.
\]
Here, $\tau$ grows more slowly than the total depleted region, indicating that only a fraction of the depletion layer contributes to electrical conduction. The characteristic length $\tau$ therefore probes a partially depleted, partially screened conducting channel rather than the full depletion width.

\paragraph*{c. Interpretation of slope ratio:}
The ratio
\[
\frac{(d\tau/d\lambda_D)}{(d\tau/dW_0)} \approx 3.44
\]
demonstrates that $\tau$ responds significantly more strongly to changes in the Debye length than to changes in the depletion width. This confirms that the apparent crossover in the Hall-extracted effective mobility is controlled primarily by long-range electrostatic screening rather than by the absolute extent of the depleted region. The persistent ordering $\lambda_D < \tau < W_0$ indicates that $\tau$ marks the depth at which surface electric fields substantially distort the electrostatic potential, but not necessarily the boundary of full depletion.

Taken together, these trends identify $\tau$ as an electrostatic conducting-thickness scale that interpolates smoothly between screening-controlled and depletion-controlled limits. Its magnitude and scaling behavior provide the physical basis for the empirical extended-exponential parameterization and for the practical design rule $t \gtrsim 3\tau$ required to preserve a near-bulk effective Hall response in thinned HPGe.

\subsection{Thickness-Dependent Effective Hall Mobility}

\begin{figure}[htbp]
    \centering
    \includegraphics[width=0.5\textwidth]{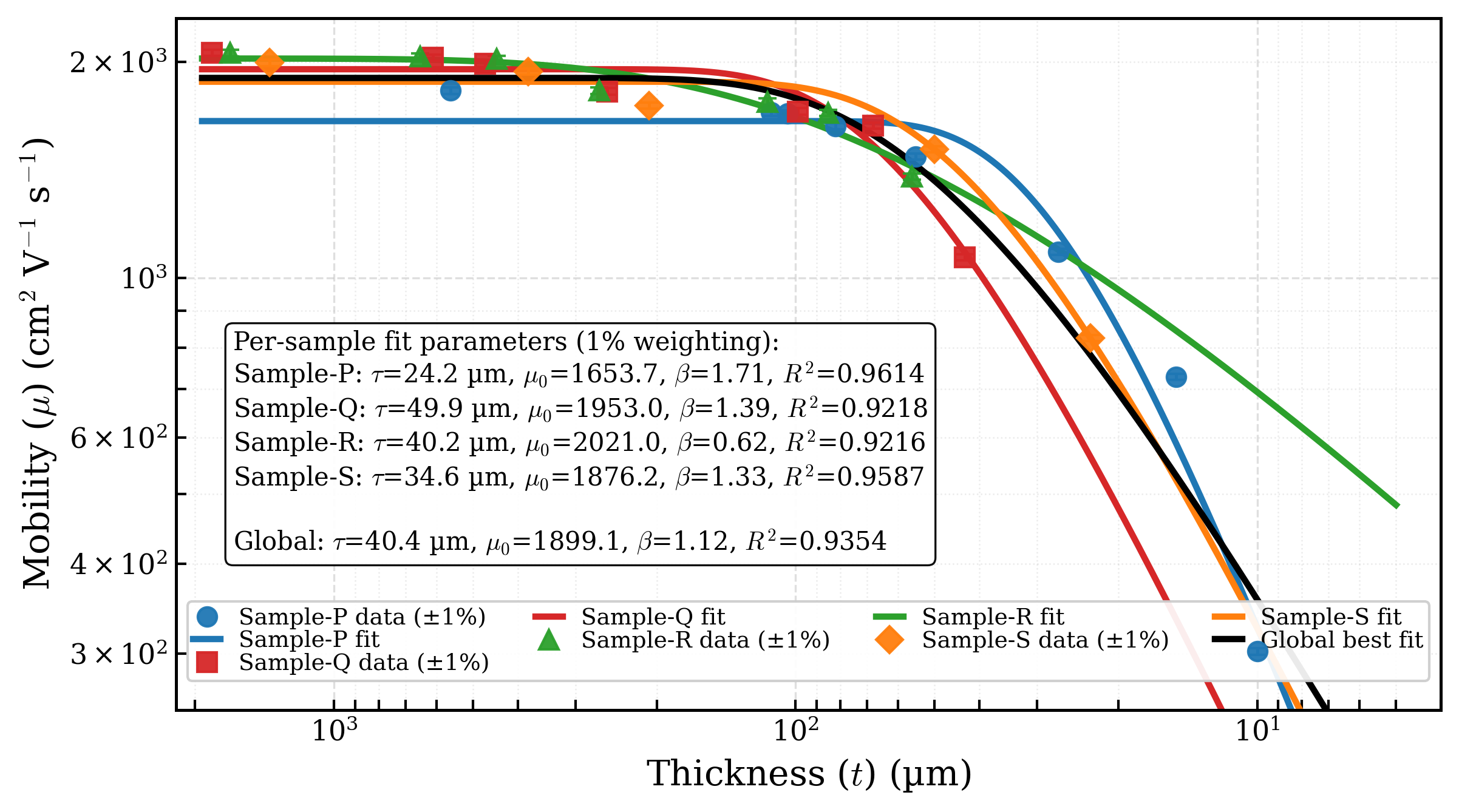}
    \caption{Effective Hall mobility for USD-grown p-type HPGe samples (P, Q, R,
    and S). Symbols denote measured data ($\pm1\%$). Colored curves show individual
    extended-exponential fits, while the black curve represents the global fit.
    All samples exhibit a smooth crossover from depletion-limited conduction at
    small thickness to a bulk-like effective Hall response as the electrically active thickness
    recovers.}
    \label{fig:mobility_hole}
\end{figure}

\begin{figure}[htbp]
    \centering
    \includegraphics[width=0.49\textwidth]{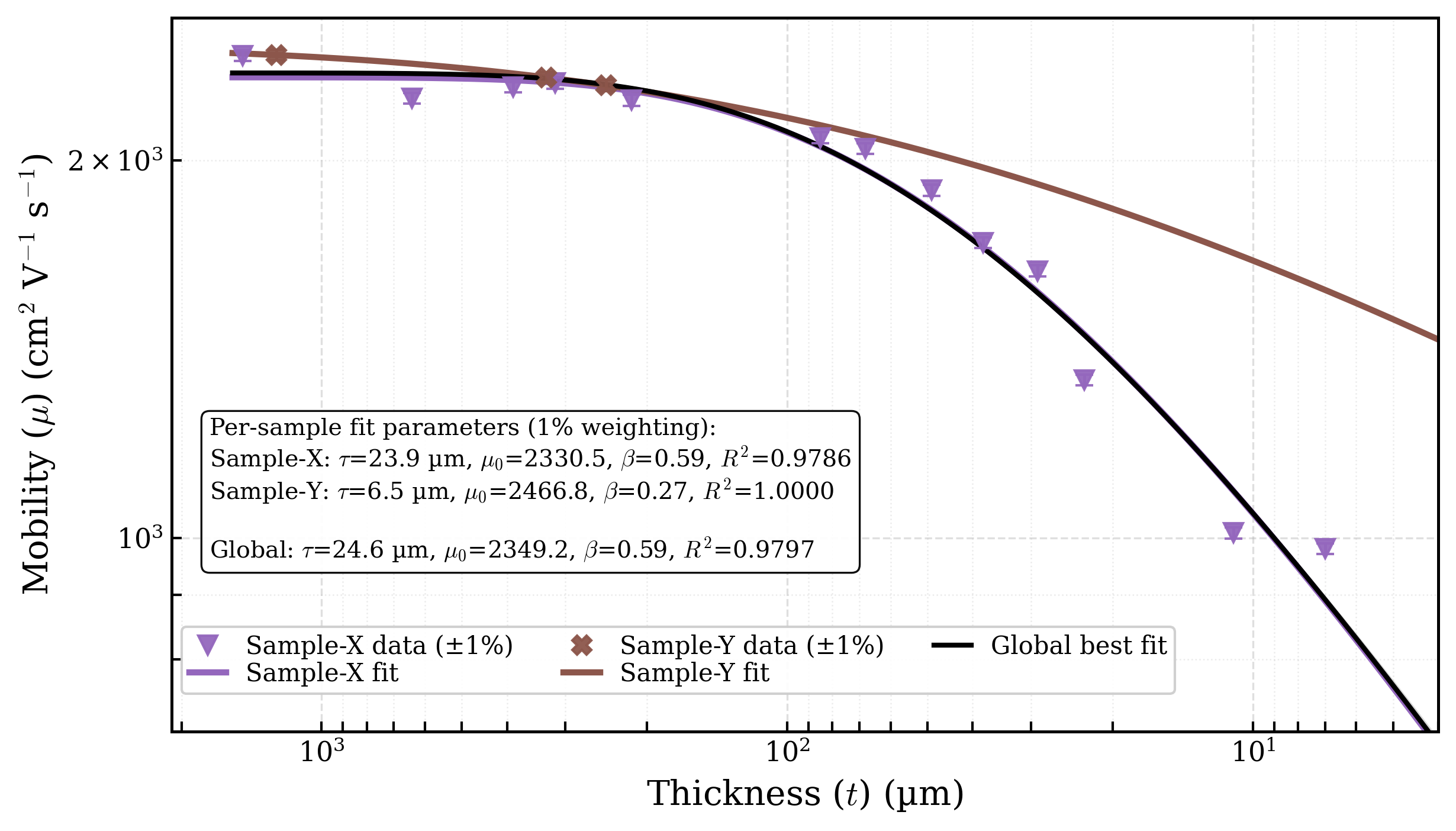}
    \caption{Effective Hall mobility for USD-grown n-type HPGe samples (X and Y).
    Symbols denote measured data, and the black curve shows the global
    extended-exponential fit.}
    \label{fig:mobility_electron}
\end{figure}

Figures~\ref{fig:mobility_hole} and~\ref{fig:mobility_electron} show the measured
effective Hall mobilities for holes and electrons as a function of the  crystal
thickness for all USD-grown samples. In the bulk regime ($t>100~\mu$m), both
carrier types asymptotically approach bulk-like room-temperature Hall mobility values
expected for high-purity HPGe,
\[
\mu_{0,h}\approx1.899\times10^3~\mathrm{cm^2/V\,.s},\qquad
\mu_{0,e}\approx2.349\times10^3~\mathrm{cm^2/V\,.s},
\]
consistent with established benchmarks reported for ultrapure Ge under Hall
measurement conditions~\cite{jacoboni1983electronic,nakano2020ultra,panth2020characterization}. In this thickness range, the electrically
active conducting region occupies nearly the full geometric thickness, and carrier
transport is dominated by phonon scattering, rendering the intrinsic bulk mobility
effectively thickness independent.

A key result of this study is that the effective Hall mobility
decreases sharply as the physical thickness is reduced below t = 100 µm. We emphasize that this behavior does not reflect
a change in the intrinsic carrier mobility. Instead, it arises from electrostatic
surface depletion, which progressively reduces the electrically active conducting
thickness as long-range surface fields, band bending, and associated space-charge
regions encroach into the bulk.

The empirical extended-exponential parameterization accurately
captures the thickness dependence of the effective Hall mobility
data for all samples with $R^2>0.92$. For holes (samples P, Q, R, and S), the global
best-fit parameters are $\mu_0=1899.1~\mathrm{cm^2/V\,.s}$,
$\tau_h=40.4~\mu$m, and $\beta_h=1.12$. For electrons (samples X and Y), the
corresponding values are $\mu_0=2349.2~\mathrm{cm^2/V\,.s}$,
$\tau_e=24.6~\mu$m, and $\beta_e=0.59$. The smaller values of $\tau_e$ and $\beta_e$
are consistent with shorter electrostatic screening lengths and reduced sensitivity of the
conduction band to electrostatic confinement, leading to a weaker thickness
dependence of the effective Hall mobility for electrons compared to holes.

Across all samples, the close correspondence between the fitted characteristic
length $\tau$ and the calculated Debye length $\lambda_D$ demonstrates that the
observed crossover in the Hall-extracted effective mobility reflects primarily electrostatic
depletion rather than interface roughness or boundary scattering. In practical
terms, maintaining thicknesses $t$ several times larger than $\tau$ ensures that
the electrically active conducting region spans most of the sample, and that the
effective Hall mobility remains within a few percent of the intrinsic bulk mobility value
expected for HPGe.
The effective Hall mobility exhibits a thickness-dependent decrease; this arises because surface depletion reduces the eletrically active conducting thickness relative to the geometric thickness used in the Hall analysis.

\subsection{Model Validation}

\begin{figure}[htbp]
    \centering
    \includegraphics[width=0.49\textwidth]{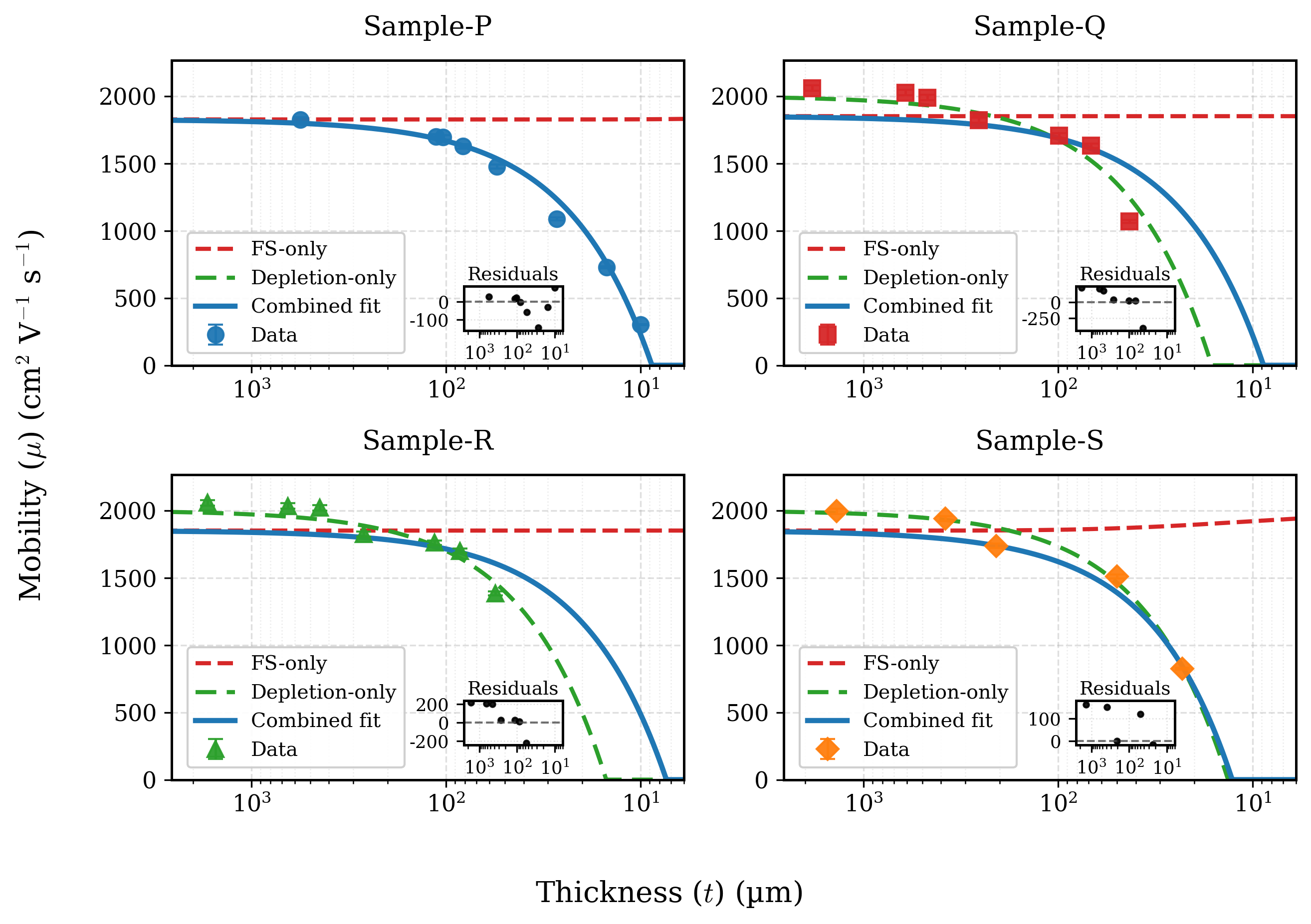}
    \caption{Model comparison for effective Hall mobility of p-type HPGe samples
    (P, Q, R, and S). The FS-only model (red dashed) predicts nearly
    thickness-independent behavior. The depletion-only model (green) reproduces
    the overall trend but fails to capture the curvature at small thickness. The
    extended-exponential parameterization (blue solid) yields small, structureless
    residuals (inset) across the full thickness range.}
    \label{fig:hole_model}
\end{figure}

\begin{figure}[htbp]
\centering
\includegraphics[width=0.49\textwidth]{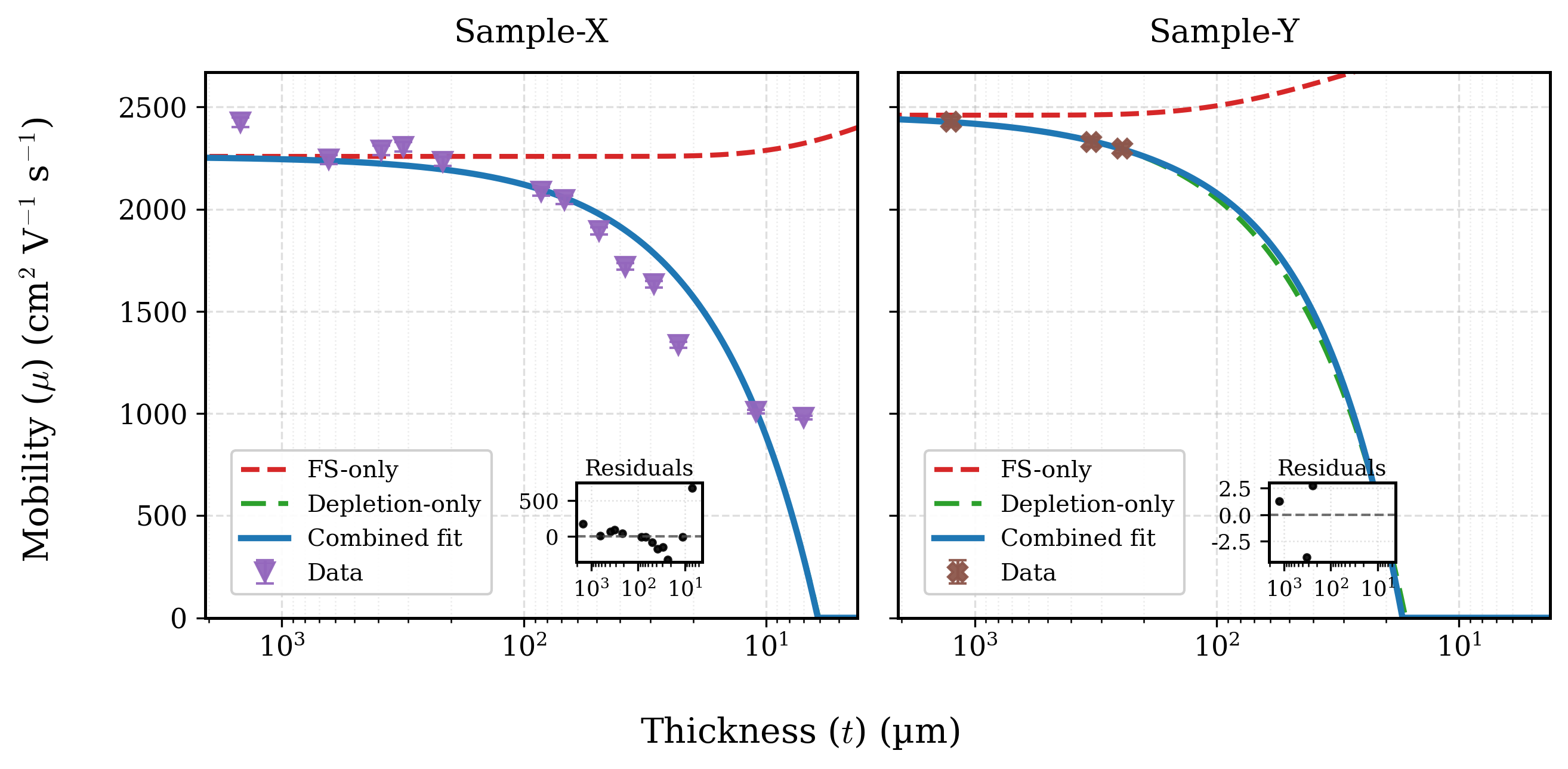}
\caption{Model comparison for effective Hall mobility of n-type HPGe samples (X
and Y). Only the extended-exponential parameterization captures both the
high-thickness saturation and the steep reduction at small thickness, with
minimal residuals (inset).}
\label{fig:electron_model}
\end{figure}

The empirical extended-exponential parameterization was benchmarked against two
classical limiting descriptions:  
(i) a Fuchs--Sondheimer (FS-only) boundary-scattering model, and  
(ii) a depletion-only electrostatic model in which the electrically active
conducting thickness is reduced by surface band bending.  
Figures~\ref{fig:hole_model} and~\ref{fig:electron_model} summarize these
comparisons for p-type and n-type samples, respectively.

For thick samples ($t\gtrsim100~\mu$m), all three descriptions converge to the
same bulk limit, consistent with an electrically active region that spans the
full geometric thickness. As the samples are thinned, the FS-only model fails to
reproduce the data: because the phonon-limited mean free path
$\lambda_{\mathrm{mfp}}$ is extremely short, the predicted correction is weak and
nearly linear, failing to account for the pronounced reduction in the
effective Hall mobility.  This
discrepancy supports the conclusion that the observed thickness dependence reflects electrostatic depletion of the conducting channel, rather than diffuse boundary scattering or any modification of the intrinsic bulk mobility.

By contrast, the extended-exponential form accurately reproduces the
Hall-extracted effective mobility across the full thickness range, with small
and non-systematic residuals. The fitted characteristic lengths $\tau_h$ and
$\tau_e$ lie between the Debye screening length $\lambda_D$ and the depletion
width $W_0$, consistent with the hierarchy in Eq.~\eqref{eq:hierarchy} and
Fig.~\ref{fig:lengths}. This agreement supports the interpretation that the
observed thickness dependence of the Hall-extracted effective mobility reflects
electrostatic depletion of the conducting channel rather than diffuse boundary
scattering or any modification of the intrinsic carrier mobility.

Overall, these comparisons demonstrate that a single compact empirical
parameterization is sufficient to describe the electrostatically controlled
crossover in the effective conducting fraction from bulk-like to
depletion-limited conduction in thinned HPGe. The parameters $\mu_0$, $\tau$,
and $\beta$ therefore provide a physically transparent and device-relevant
description of how electrostatic geometry, rather than microscopic scattering
mechanisms, constrains the Hall-extracted effective mobility in ultrathin
Ge-based detectors and electronic or quantum devices.

\section{Discussion}

It is important to emphasize that the electrostatic length scales, mobility
behavior, and effective Hall mobility trends discussed here are strictly
validated at room temperature ($T = 295$~K), where phonon scattering dominates
and the carrier mean free path remains short. All conclusions drawn in this
section are therefore limited to this temperature regime.

\medskip
\noindent
\subsection{Dimensionality and interpretation of Hall mobility}
We first clarify the distinction between the present system and true
quasi-two-dimensional electron gases (2DEGs) realized in GaAs/AlGaAs,
Si/SiGe, or Si/SiO$_2$ heterostructures. Over the full thickness range
investigated here (7~$\mu$m–2.7~mm), the mechanically thinned HPGe samples do
not exhibit quantum confinement or sub-band quantization, and charge transport
remains three-dimensional within the electrically neutral region. The mobility
reported in this work is therefore not a 2DEG mobility in the quantum-well
sense, but an effective Hall mobility extracted using a Van der Pauw
formalism applied to a finite-thickness three-dimensional conductor. In the
thick-sample limit, where the electrically active thickness approaches the
geometric thickness, the Hall mobility directly reflects the intrinsic bulk
mobility and thus remains a valid metric of crystal quality. In contrast, in
the thinned regime the Hall-extracted mobility is progressively reduced by
electrostatic depletion of the conducting channel rather than by changes in
microscopic scattering mechanisms.

\medskip
The thickness dependence observed in the present measurements reflects a clear
transition in the effective Hall mobility of high-purity Ge from a
bulk-like regime to a depletion-limited regime at $T = 295$~K. In the
thick-sample limit ($t \gtrsim 100~\mu$m), the electrically active conducting
region spans nearly the full geometric thickness, and the extracted Hall
mobilities,
$\mu_{0,h}\approx1.899\times10^{3}$~cm$^{2}$/V·s and
$\mu_{0,e}\approx2.349\times10^{3}$~cm$^{2}$/V·s,
are consistent with bulk-like Hall mobilities reported for high-purity Ge
\cite{jacoboni1983electronic,fischetti1996band,panth2020characterization}. In
this regime, carrier momentum relaxation is dominated by acoustic and optical
phonon scattering, and the intrinsic carrier mobility is effectively
independent of thickness.

As the physical thickness is reduced below $t\approx100~\mu$m, the extracted
Hall mobility decreases sharply. We emphasize that this behavior does not imply
a modification of intrinsic carrier mobility. Instead, it arises from
electrostatic surface depletion, which progressively reduces the electrically
active conducting thickness as long-range surface electric fields and band
bending encroach into the bulk. The observed crossover therefore reflects a
geometric and electrostatic constraint on current flow rather than a transition
to a new microscopic transport regime.

Classical diffuse boundary scattering cannot account for this behavior. The
phonon-limited mean free path,
$\lambda_{\mathrm{mfp}}\approx44$~nm, remains more than three orders of
magnitude smaller than the thinnest samples investigated. Consequently, all
specimens lie well outside the geometric-confinement regime in which the
Fuchs–Sondheimer (FS) formalism is applicable. This explains the failure of the
FS-only prediction in
Figs.~\ref{fig:mobility_hole}–\ref{fig:electron_model}, which yields only weak,
nearly thickness-independent corrections to the effective Hall mobility.

By contrast, the characteristic electrostatic lengths extracted from the
extended-exponential parameterization,
$\tau_h\approx40.4~\mu$m and $\tau_e\approx24.6~\mu$m, fall between the Debye
screening length $\lambda_D$ and the depletion width $W_0$, satisfying the
hierarchy
\[
\lambda_{\mathrm{mfp}} \ll \lambda_D < \tau \lesssim W_0 .
\]
This ordering directly links the observed reduction in effective Hall mobility
to long-range electrostatic surface fields, which compress the neutral
conducting channel without altering local scattering mechanisms. The linear
trends in Fig.~\ref{fig:lengths} further quantify this relationship: $\tau$
increases with $\lambda_D$ with a slope exceeding unity, indicating that the
recovery of bulk-like transport with increasing thickness is strongly governed
by electrostatic screening, while its weaker dependence on $W_0$ reflects the
partial, but not complete, contribution of the depletion region to electrical
conduction.

The fitted exponents ($\beta_h=1.12$, $\beta_e=0.59$) provide a coarse-grained
description of how band structure and electrostatics jointly influence the
thickness dependence of the effective Hall mobility. The sublinear $\beta_e$
reflects the comparatively weak anisotropy and nonparabolicity of the Ge
conduction band, while $\beta_h>1$ is consistent with the strongly warped
valence band and its enhanced sensitivity to electrostatic confinement
\cite{luttinger1956quantum,bir1974symmetry}. Although these parameters are
extracted from micrometer-scale samples—where quantum confinement and sub-band
formation are absent—the pair $(\tau,\beta)$ captures, at a phenomenological
level, how electrostatic geometry modulates the electrically active conducting
thickness at room temperature.

\subsection{Comparative Advantages of
Thinned Bulk HPGe}

To provide a broader context for
the present results, Table II compares the dimensionality, operating temperature, and representative mobilities of established Ge- and Si-based transport
platforms. For bulk Ge, literature values typically report intrinsic, lattice limited mobilities at low field. In contrast, the values reported in this work
represent the bulk-limit mobility, $\mu_0$
extracted from the thick-sample regime.
It is important to distinguish these from the effective Hall mobilities observed at smaller thicknesses, where electrostatic depletion reduces the conducting
cross-section. We further note that the bulk-limit mobilities extracted here are
specific to the measured carrier density ($N_{\mathrm{eff}}$ ) and compensation levels of the
detector-grade material utilized; thus, they are not intended to represent the
theoretical maximum for idealized, high-purity Ge. Unlike Ge-on-insulator
(GeOI) or heterostructure-based 2DEG systems, our approach maintains three dimensional bulk transport while systematically probing the micrometer regime.
The primary advantage of this platform lies in its preservation of near-intrinsic
bulk properties at extremely low impurity concentrations. By avoiding heterointerfaces, strain engineering, and gate oxides, we eliminate significant sources of
interfacial disorder, trapped charge, and 1/f noise. This makes thinned, bulk-grown HPGe an ideal candidate for radiation detectors and phonon-mediated
quantum sensors, where carrier lifetime and electrostatic stability are prioritized
over record-high sheet mobilities.

\begin{table*}[t]
\centering
\fontsize{8}{12}\selectfont
\caption{Comparison of representative carrier mobility in Ge- and Si-based material systems at or near room temperature.}
\label{tab:comparison}

\begin{ruledtabular}
\begin{tabular}{lcccccc}
System &
Dim. &
$T$ (K) &
Mobility (cm$^{2}$/V$\cdot$s) &
Carrier density &
Thickness / Confinement \\
\hline

Bulk HPGe (this work) &
3D &
295 &
$\mu_{0,h}\!\approx\!1.9\times10^{3}$,
$\mu_{0,e}\!\approx\!2.3\times10^{3}$ &
$10^{10}$--$10^{11}$~cm$^{-3}$ &
7--2700~$\mu$m \\

Bulk Ge (literature)~\cite{intrinsic_mobility} &
3D &
300 &
$\mu_h \sim 1900$,
$\mu_e \sim 3900$ &
$\sim10^{13}$--$10^{14}$~cm$^{-3}$ &
Bulk \\

Ge-on-Si epitaxial film (strain-engineered)~\cite{wei2020jcg} &
3D &
300 &
$\mu_h \approx 1.25\times10^{3}$ &
$\sim10^{16}$~cm$^{-3}$ &
$\sim$500~nm \\

Strained Ge/SiGe quantum well~\cite{Myronov2014JJAP} &
2D &
300 &
$\mu_h^{2D} \approx 4.0\times10^{3}$ &
$p_{2D} \sim 10^{11}$~cm$^{-2}$ &
$\sim$20~nm \\

Si/SiO$_2$ MOS inversion layer~\cite{Takagi1994TED} &
2D &
300 &
$\mu_e \sim (1$--$3)\times10^{3}$,
$\mu_h \sim (2$--$8)\times10^{2}$ &
$10^{11}$--$10^{12}$~cm$^{-2}$ &
Electrostatic inversion layer \\

\end{tabular}
\end{ruledtabular}

\normalsize
\end{table*}

\subsection{Extrapolation toward the nanoscale limit}

Although chemical--mechanical thinning cannot reliably produce nanometer-thick
HPGe layers with atomically smooth surfaces, the experimentally established
electrostatic trends enable a physically grounded extrapolation toward the
nanoscale. The strong correlation between the characteristic electrostatic
length $\tau$ and the Debye screening length $\lambda_D$ for lightly doped
samples (P, Q, R, and S) indicates that reducing the physical thickness into the
few--tens-of-nanometers range would place the entire structure within the
electrostatic screening region. In this limit, the electrically active
conducting thickness is expected to be strongly constrained by the surface
potential profile, leading to a rapid reduction of the effective Hall
mobility as $t$ approaches $\lambda_D$, even though the intrinsic bulk mobility
remains unchanged.

For the more heavily doped samples (X and Y), the weaker scaling of $\tau$ with
the depletion width $W_0$ suggests that the conducting channel is already
significantly compressed by surface space-charge fields at micrometer
thicknesses. As the physical thickness is further reduced toward the phonon-
limited mean free path ($t\sim\lambda_{\mathrm{mfp}}$), surface-scattering
mechanisms---which are negligible in the present micrometer-scale samples---are
expected to become increasingly important. In this ultrathin limit, transport
would gradually transition from an electrostatically constrained regime to a
boundary-influenced regime, consistent with observations in ultrathin Ge films
and Ge-on-insulator structures~\cite{nakatsuka2019enhanced,imajo2021strain}.

In this sense, the present results bridge two complementary thickness regimes:
micrometer-scale bulk-grown HPGe, where electrostatic depletion governs the
electrically active conducting thickness, and the nanoscale limit, where both
electrostatic confinement and surface scattering jointly influence the
effective Hall mobility. A detailed treatment of cryogenic temperatures, where
carrier mean free paths increase and additional scattering and decoherence
mechanisms may emerge, is beyond the scope of this work and will be addressed
in future studies.

\subsection{Device implications}

The present results have direct implications for the design of HPGe radiation
detectors, thin-body Ge electronics, and emerging quantum devices. For
low-threshold radiation detectors, reducing the physical thickness below a few
times the characteristic electrostatic length $\tau$ leads to a pronounced
reduction in the effective Hall mobility due to electrostatic depletion
of the conducting channel. This reduction increases carrier transit times and
can elevate electronic noise, even though the intrinsic bulk mobility remains
unchanged. Maintaining active regions in the regime $t \gtrsim 3\tau$ preserves
near-bulk transport, whereas devices with $t \lesssim \tau$ operate in a
depletion-controlled regime with strongly reduced electrically active
conduction. In practice, these effects can be mitigated by maintaining critical
drift regions thicker than the relevant electrostatic length scales or by
reducing surface electric fields through appropriate passivation, dielectric
engineering, and optimized contact geometries.

For nanoscale Ge-based platforms, including phonon-coupled quantum sensors and
Ge-based qubit architectures, the results demonstrate that impurity control
alone is insufficient to preserve favorable transport properties. Precise
engineering of the surface potential and depletion geometry becomes essential
as dimensions are reduced, since electrostatic confinement can severely limit
the electrically active conducting thickness well before classical boundary
scattering becomes relevant. The parameter set $(\mu_0,\tau,\beta)$ extracted
here therefore provides a compact, device-level framework for estimating
electrostatic mobility penalties in thinned Ge structures under realistic
operating conditions.

In summary, this work establishes a quantitative and physically grounded
framework linking Debye screening, surface depletion geometry, and the
thickness-dependent effective Hall mobility of HPGe at room
temperature. By explicitly connecting the parameters of the empirical extended exponential parameterization to the observed transport trends, we
clarify the transition from bulk-like phonon-limited transport to
depletion-limited conduction. These insights provide practical guidance for the
optimization of Ge-based detectors and thin-body semiconductor devices that
operate at the boundary between bulk and mesoscale physics, and they establish a
foundation for future extensions to cryogenic operation and quantum device
architectures.

\section{Conclusion}

We have experimentally quantified how the effective Hall mobility in
high-purity germanium evolves with sample thickness over a wide micrometer-scale
range at room temperature. In the bulk limit, the extracted Hall mobilities
approach bulk-like values consistent with those reported for high-purity Ge,
confirming that carrier transport is phonon-dominated and that the intrinsic
bulk mobility is independent of thickness in the regime studied. As the crystals
are thinned, the measured Hall mobility exhibits a pronounced reduction that
cannot be explained by classical diffuse boundary scattering, since the
phonon-limited mean free path remains orders of magnitude smaller than the
physical thickness for all samples.

Instead, the observed behavior is governed by electrostatic surface depletion.
The characteristic lengths extracted from the empirical extended-exponential
parameterization correlate with the expected electrostatic scales with this relation
\[
\lambda_{\mathrm{mfp}} \ll \lambda_D < \tau \lesssim W_0,
\]
demonstrating that the reduction in effective Hall mobility is governed by
Debye screening and surface depletion rather than by geometric confinement or
a modification of microscopic scattering mechanisms. The strong correlation
between $\tau$ and the Debye length in lightly doped samples, together with the
weaker dependence on the depletion width in more heavily doped material,
quantifies how the electrically active conducting thickness evolves with doping
and surface band bending. These trends confirm that the apparent mobility
crossover reflects the penetration depth of electrostatic perturbations into
the crystal, not interface roughness or boundary scattering.

Although the present measurements are confined to the micrometer regime, the established hierarchy of length scales provides a physically grounded framework for discussing the scaling of transport properties toward the nanoscale. As thickness approaches tens of nanometers, the electrically active volume is expected to become fully confined
within the electrostatic screening region, with transport progressively
transitioning from purely depletion-limited behavior to a regime where surface
scattering also contributes, consistent with observations in ultrathin epitaxial
and Ge-on-insulator films. It should be emphasized that all conclusions drawn in this study are strictly limited to room temperature transport ($T = 295\,\mathrm{K}$). At cryogenic temperatures, the underlying
transport physics transitions significantly: carrier freeze-out, modified screening lengths, and drastically increased mean free paths may alter the relative
dominance of electrostatic depletion versus surface scattering. Consequently,
extending this empirical framework to the cryogenic regime represents a distinct and important direction for future work.

Overall, this study establishes a robust, quantitative framework
that integrates Debye screening and surface depletion to account for the
thickness-dependent effective Hall mobility at room-temperature HPGe. By explicitly
connecting the parameters $(\mu_0,\tau,\beta)$ of the empirical
extended-exponential description to fundamental electrostatic length scales, we
provide a physically transparent interpretation of transport from bulk-like to
depletion-limited regimes. A practical design rule follows directly: maintaining
$t \gtrsim 3\tau$ preserves near-bulk transport, whereas structures with
$t \lesssim \tau$ operates in a depletion-controlled regime with strongly reduced
electrically active conduction. These insights inform the optimization of
Ge-based radiation detectors, thin-body electronics, and emerging quantum devices
that must retain high transport performance under geometric confinement.

\section*{Author Contributions}
N.~Budhathoki performed sample fabrication, surface processing,
Hall-effect measurements, data analysis, and manuscript preparation.
D.-M.~Mei provided conceptual guidance, supervision, and theoretical
interpretation.
S.~Bhattarai, S.~Chhetri, K.~Dong, S.~Panamaldeniya, A.~Prem, and A.~Warren 
contributed to crystal growth and material preparation.

\section*{Acknowledgments}

This work was supported in part by the National Science Foundation (NSF)  under 
Grants No.~NSF OISE-1743790, NSF PHYS 2117774, NSF OIA 2427805, NSF PHYS-2310027, and NSF OIA-2437416, and by the U.S. 
Department of Energy (DOE) under Grants No.~DE-SC0024519 and DE-SC0004768. 
Additional support was provided by a research center funded by the State 
of South Dakota.

\section*{Data Availability}

The data that support the findings of this study are available from the 
corresponding author upon reasonable request.

\bibliographystyle{apsrev4-2}
\bibliography{name1}

\end{document}